\documentclass[aps,preprintnumbers,prb,showpacs,superscriptaddress]{revtex4}
\usepackage{latexsym}
\usepackage{amssymb}
\usepackage{graphicx}
\begin{document}
\title{Vanishing spin-Hall current in a diffusive Rashba two-dimensional
electron system: A quantum Boltzmann equation approach}
\author{S. Y. Liu}
\email{liusy@mail.sjtu.edu.cn}
\affiliation{Department of Physics, Shanghai Jiaotong University, 1954
Huashan Road, Shanghai 200030, China}
\affiliation{Department of Physics and Engineering Physics, Stevens
Institute of Technology, Hoboken, New Jersey 07030, USA}
\author{X. L. Lei}
\affiliation{Department of Physics, Shanghai Jiaotong University, 1954
Huashan Road, Shanghai 200030, China}
\author{ Norman J. M. Horing}
\affiliation{Department of Physics and Engineering Physics, Stevens
Institute of Technology, Hoboken, New Jersey 07030, USA}

\date{\today}
\begin{abstract}
We present a quantum Boltzmann equation analysis of the spin-Hall
effect in a diffusive Rashba two-dimensional electron system.
Within the framework of the self-consistent Born approximation, we
consider the roles of disorder-induced quasiclassical relaxation,
collisional broadening of the quasiparticles, and the
intracollisional field effect in regard to spin-Hall dynamics. We
present an analytical proof that the spin-Hall current vanishes,
independently of the coupling strength, of the quasiparticle
broadening, of temperature and of the specific form of the
isotropic scattering potential. A sum relation of the collision
terms in a helicity basis is also examined.
\end{abstract}

\pacs{72.10.-d, 72.25.Dc, 73.50.Bk}
\maketitle
\section{Introduction}
The spin-Hall effect refers to the appearance of a net polarized
spin flow along the direction perpendicular to an external applied
dc electric field. This phenomenon suggests a possible mechanism
to control spin dynamics using an electric field and may be
employed to resolve the challenge of spin injection in the
emerging field of spintronics. In early studies, the spin-Hall
effect was associated with the disorder-induced spin-orbit (SO)
interaction, of an extrinsic nature.\cite{DP,HS} More recently,
another disorder-free spin-Hall effect was predicted,
respectively, in bulk $p$-doped semiconductors\cite{Zhang} and in
Rashba two-dimensional (2D) systems.\cite{Sinova} This intrinsic
effect essentially arises from disorder-independent SO
interactions, such as the Rashba SO coupling, the SO interaction
involved in the Luttinger model, {\it etc.} Experimentally, the spin-Hall
effect has been observed in a bulk $n$-type semiconductor\cite{Kato} and
in a 2D heavy-hole system.\cite{Wunderlich}

In Rashba 2D electron systems, the spin-Hall current can be
strongly affected by a spin-conserving interaction between
electrons and impurities. This issue has been investigated in the
diffusive regime extensively by various methods, including the
Kubo formula,\cite{Loss1,Nomura,Loss2,Dimitrova, Bauer, Raimondi}
the spin-density matrix method,\cite{Khaetskii} and a
nonequilibrium Green's function
approach.\cite{Halperin,Liu1,Nagaosa} Physically, to examine the
effect of disorder on spin-Hall current, we must simultaneously
take into account collisional broadening and quasiclassical
relaxation, as well as the intracollisional field
effect\cite{Horing, Khan, Regiani} produced by the action of the
electric field during the course of electron-impurity scattering.
The most important disorder effect on spin-Hall current is
associated with quasiclassical relaxation. This scattering process
yields an additional term to the spin-Hall current, causing it to
vanish.\cite{Loss2,Dimitrova, Bauer,
Raimondi,Khaetskii,Halperin,Liu1} In addition, the
collision-induced spread of the quasiparticle density of states
(DOS)-namely, collisional broadening-also influences the spin-Hall
effect. It reduces the value of the intrinsic spin-Hall
conductivity with increasing broadening of the DOS.\cite{Loss1,
Nomura} For short-range disorder, the total spin-Hall current
still vanishes when collisional broadening in both the intrinsic
and quasiclassical terms of the spin-Hall current is
considered.\cite{Loss2,Dimitrova,Raimondi,Bauer} There is yet
another disorder-induced quantum effect-namely, the
intracollisional field effect-which can not be
neglected.\cite{Horing, Khan, Regiani} During the scattering
process, the DOS of the quasiparticles can be modified by the
external electric field, whereupon relaxation becomes dependent on
this field even in linear response. To date, only Sugimoto {\it et
al.} have tried to investigate the spin-Hall effect including
these three processes simultaneously.\cite{Nagaosa} Performing a
numerical calculation of the Keldysh nonequilibrium Green's
function in a spin basis, they found a generally nonvanishing
spin-Hall current: the value of the spin-Hall conductivity they
obtained depends on the spin-orbit coupling constant, the
transport lifetime, and the Fermi momentum.

In this paper, we analytically investigate the spin-Hall effect in
Rashba 2D systems using the nonequilibrium Green's function
method, essentially the same as that used by Sugimoto {\it et
al.},\cite{Nagaosa} but in a helicity basis. We carefully analyze
the nonequilibrium retarded and lesser Green's functions by
simultaneously taking account of quasiclassical relaxation,
collisional broadening, and the intracollisional field effect. We
prove analytically that the spin-Hall current vanishes,
irrespective of the spin-orbit coupling strength, of the form of
the isotropic impurity potential, of temperature, and of the
collisional broadening of the DOS. We also discuss a sum relation
of the relaxation terms in the helicity basis.

This paper is organized as follows. In Sec. II, we present the
kinetic equation for the "lesser" Green's function. The vanishing
of spin-Hall current is proven in Sec. III. Finally, we review our
results in Sec. IV. Several appendixes provide requisite details
which could be read before Eq.\,(\ref{DYSS2}) by readers desiring
a "proof first" exposition.

\section{Kinetic equation for a 2D Rashba system}

We consider a 2D Rashba electron system with a single-particle
noninteracting Hamiltonian given by\cite{Rashba}
\begin{equation}
{\check h}_0({\bf p})=\frac{{ p}^2}{2m}+\alpha {\bf p}\cdot
({\bf n}\times {\bf \sigma}).\label{ham}
\end{equation}
Here, ${\bf p}\equiv (p_x,p_y)\equiv (p\cos\phi_{\bf
p},p\sin\phi_{\bf p})$ is the 2D electron momentum, $m$ is the
effective mass, ${\bf \sigma}\equiv (\sigma_x,\sigma_y,\sigma_z)$
are the Pauli matrices, $\alpha$ is a spin-orbit coupling
constant, and ${\bf n}$ is a unit vector perpendicular to the 2D
electron plane. By a local unitary spinor transformation $ U_{\bf
p}$,
\begin{equation}
U_{\bf p}=\frac 1 {\sqrt{2}}\left (
\begin{array}{cc}
1&1\\
i{\rm e}^{i\phi_{\bf p}}&-i{\rm e}^{i\phi_{\bf p}}
\end{array}
\right ),\label{Uni}
\end{equation}
the Hamiltonian (\ref{ham}) can be diagonalized as ${\hat
h}_0({p}) = U^+_{\bf p}\check h_0({\bf p}) U_{\bf p}={\rm
diag}(\varepsilon_{1}(p), \varepsilon_{2}({p}))$ with
$\varepsilon_\mu ({p})=\frac{{p}^2}{2m}+(-1)^\mu \alpha p$
($\mu=1,2$) as dispersion relations of two spin-orbit-coupled
bands. In this paper, we investigate the spin-Hall effect in a 2D
Rashba electron system driven by a constant, uniform electric
field ${\bf E}$ along the $x$ axis. In the Coulomb gauge, the
electric field can be described by a scalar potential, $V\equiv
-e{\bf E}\cdot {\bf r}$, with ${\bf r}$ as the electron
coordinate.

In a realistic 2D system, electrons experience scattering by
impurities. We assume that the interaction between electrons and
impurities can be characterized by an isotropic potential,
$V(|{\bf p}-{\bf k}|)$, corresponding to scattering of an electron
from state ${\bf p}$ to state ${\bf k}$.

The nonequilibrium Green's functions $\check {\rm G}^{r,<}_{\bf
p}$ in spin basis for 2D electron systems with Rashba SO coupling
are defined as usual. They are $2\times 2$ matrices and obey Dyson
equations presented in Appendix A. For brevity, hereafter, we
employ a subscript ${\bf p}$ to denote the arguments of the
Green's functions and self-energies, $({\bf p},\omega)$. It is
most convenient to study the lesser Green's function in the
helicity basis, $\hat {\rm G}^{<}_{\bf p}= U_{\bf p}^+ \check {\rm
G}^{<}_{\bf p}  U_{\bf p}$, in which the unperturbed lesser
Green's function is diagonal. Using the transformation from spin
basis to helicity basis, the Dyson equation for $\hat {\rm
G}^<_{\bf p}$ can be written as
\begin{widetext}
\begin{equation}
ie{\bf E}\cdot \left (\frac{\partial}{\partial {\bf p}} +{\bf
p}\frac{\partial}{\partial \omega}\right )\hat {\rm G}^<_{\bf p}+
ie{\bf E}\cdot [\hat {\rm G}^<_{\bf p},\nabla_{\bf p}U^+_{\bf p}U_{\bf
p}] +\alpha p[\sigma_z,\hat {\rm G}^<_{\bf p}]+\frac{i\alpha}{2}
e{\bf E}\cdot \frac{\partial}{\partial \omega} \hat {\bf B}^<_{\bf
p}= \hat I_{l},\label{DYSS2}
\end{equation}
\end{widetext}
with $\hat {\bf A}^{<}_{\bf p}=[{\bf N}_{\bf p}, \hat {\rm G}_{\bf
p}^{<}]$, $\hat {\bf B}^{<}_{\bf p}=\{{\bf N}_{\bf p}, \hat {\rm
G}_{\bf p}^{<}\}$, ${\bf N}_{\bf p}=U^+_{\bf p}({\bf n}\times {\bf
\sigma} )  U_{\bf p}$, and $\hat I_{l}$ as a scattering term. In
general, $\hat I_{l}$ has a complicated form because the momentum
and time variables of the Green's functions and self-energies are
intertwined due to the presence of the electric field.\cite{Jauho}
Physically, this feature involves the intracollisional field
effect\cite{Horing, Khan, Regiani} and results in an additional
electric-field-dependent scattering. In this paper, we restrict
our considerations to the linear response regime. Based on this,
the scattering term can be expressed as the sum of three terms:
$\hat I_{l}=\hat I_{l1}+\hat I_{l2}+\hat I_{l3}$. The first term,
$\hat I_{l1}$ does not involve the explicit appearance of the
electric field (although it is implicit in $\hat{\rm G}^{r,<}_{\bf
p}$ and self-energies $\hat {\Sigma}^{r,<}_{\bf p}$). The terms
$\hat I_{l2}$ and $\hat I_{l3}$ arise from the first-order
gradient expansion involving explicit linear dependence on the
electric field, and they are a manifestation of the
intracollisional field effect. The explicit forms of quantities
$\hat I_{l1}$ and $\hat I_{l2}$ are similar to $\check I_{l1}$ and
$\check I_{l2}$, respectively, but with helicity-basis Green's
functions and self-energies in place of the corresponding
spin-basis ones (the forms of $\check I_{l1}$ and $\check I_{l2}$
are provided in Appendix A). The term $\hat I_{l3}$ arises from
the local nature of the transformation and its form is obtained
from $\hat I_{l2}$ by using $\frac i 2\nabla_{\bf p}\phi_{\bf
p}[\cdots,\sigma_x]$ instead of operator $\nabla_{\bf p}$.

From the explicit form of the scattering term, it may be seen that
$\hat I_{l}$ relates to the retarded Green's function ${\hat {\rm
G}}_{\bf p}^r$ in which the collisional broadening is considered.
Detailed properties of ${\hat {\rm G}}_{\bf p}^r$ are analyzed in
Appendix B. We find that the linear electric field correction to
the retarded Green's function $\hat {\rm G}^r_{1{\bf p}}$ in a 2D
Rashba system does {\it not} vanish in general (although it has
vanishing diagonal elements). Also, we make clear that disorder
scattering is not essential for the nonvanishing of $\hat {\rm
G}^r_{1{\bf p}}$: $\hat {\rm G}^r_{1{\bf p}}$ exists even when the
electron-impurity collisions are ignored.

It is well known that the linear electric field correction to the
retarded Green's function vanishes for a one-band electron
gas.\cite{Mahan2} However, for two-band systems, if a transition
between the two bands is present, $\hat {\rm G}^r_{1{\bf p}}$ no
longer vanishes (see, for example, Ref.\,\onlinecite{Jauho} p.
215). In the 2D Rashba system that we study, there are two
spin-orbit-coupled bands arising from the structure asymmetry and
a polarization process can be induced between these two bands by a
dc field.\cite{Liu2} Accordingly, the linear dc electric field
correction to $\hat {\rm G}_{\bf p}^r$ in a 2D Rashba electron
system is nonvanishing.

In Eq.\,(\ref{DYSS2}), the electron-impurity scattering is
embedded in the self-energies $\hat{\Sigma}^{r,<}_{\bf p}$. In the
present paper, we consider the disorder collisions only in the
self-consistent Born approximation. It is widely accepted that
this treatment is sufficiently accurate to analyze transport
properties in the diffusive regime. On this basis, the
self-energies in helicity basis take the forms
\begin{equation}
\hat{\Sigma}^{r,<}_{\bf p}=n_i\sum_{{\bf k}}\hat T({\bf
p},{\bf k})\hat{\rm G}^{r,<}_{\bf k}\hat T^+({\bf p},{\bf
k}),\label{SE}
\end{equation}
where $\hat T ({\bf p},{\bf k})=U^+({\bf k})V({\bf p}-{\bf k})U({\bf p})$
and $n_i$ is the impurity density.

Further, we assume that the applied dc field is weak enough that only the
linear response is required to be considered. After linearizing
Eq.\,(\ref{DYSS2}), we can write the kinetic equation for the linear electric
field correction to the lesser Green's function, $\hat {\rm
G}^<_{1{\bf p}}=\hat {\rm G}^<_{\bf p} -\hat {\rm G}^<_{0{\bf
p}}$, as
\begin{equation}
-\alpha p \hat{C}_{1}+ ie{\rm \bf E}\cdot  {\bf \nabla}_{\bf
p}\hat{\rm G}_{0{\bf p}}^< -\frac{1}{2} e {\bf \rm E} \cdot{\bf
\nabla}_{\bf p}\phi_{\bf p} \hat{D}_0+R=\hat I^{(1)}_l,\label{KE}
\end{equation}
where the matrices $\hat{C}_1$ and $\hat{D}_0$ are given by
\begin{equation}
\hat{C}_1=\left (
\begin{array}{cc}
0&-2 (\hat{\rm G}^<_{1{\bf p}})_{12}\\
2 (\hat{\rm G}^<_{1{\bf p}})_{21}&0\\
\end{array}
\right ),
\end{equation}
\begin{equation}
\hat{D}_0=\left (
\begin{array}{cc}
0&(\hat{\rm G}^<_{0{\bf p}})_{11}-(\hat{\rm G}^<_{0{\bf p}})_{22}\\
(\hat{\rm G}^<_{0{\bf p}})_{22}-(\hat{\rm G}^<_{0{\bf p}})_{11}&0
\end{array}
\right ).\label{EG}
\end{equation}
$R$ is a remainder
term which can be expressed as a total derivative with respect to
$\omega$.
$\hat {\rm G}_{0{\bf p}}^<$ is the unperturbed equilibrium lesser Green's
function. It is a diagonal matrix and simply relates to the retarded
$\hat {\rm G}^r_{0{\bf p}}$ by the Kubo-Martin-Schwinger relation\cite{KMS}
\begin{equation}
\hat{\rm G}^<_{0{\bf p}}=-2in_{\rm F}(\omega){\rm
Im}\hat{\rm G}^r_{0{\bf p}},
\end{equation}
with $n_{\rm F}(\omega)$ as the Fermi function.

The scattering term $\hat I^{(1)}_l$ is the sum of $\hat
I_{l1}^{(1)}$, $\hat I_{l2}^{(1)}$, and $\hat I_{l3}^{(1)}$, the
linear electric field parts of $\hat I_{l1}$, $\hat I_{l2}$ and
$\hat I_{l3}$. The explicit forms of these quantities are
presented in Appendix C. It should be noted that $\hat
I_{l2}^{(1)}$ and $I_{l3}^{(1)}$ are diagonal and off-diagonal
matrices, respectively. In deriving Eq.\,(\ref{KE}), the
properties of the retarded Green's function in the presence of
collisional broadening are employed: its linear electric field
correction is an off-diagonal matrix, while the unperturbed one is
diagonal.

From Eq.\,(\ref{KE}) we can formally express
the off-diagonal element of $\hat {\rm G}^<_{1{\bf p}}$ as
\begin{eqnarray}
(\hat {\rm G}_{1{\bf p}}^<)_{12}&=&\frac 1 {2\alpha p}\left \{\frac 12 e{\bf
E}\cdot \nabla_{\bf p}\phi_{\bf p}[(\hat {\rm G}_{0{\bf
p}}^<)_{11}
-(\hat {\rm G}_{0{\bf p}}^<)_{22}]\right.\nonumber\\
&&\left . +(I_l^{(1)})_{12}-R_{12}\right \}.\label{EQ12}
\end{eqnarray}
At the same time, the diagonal elements of $I_l^{(1)}$ take the
form,
\begin{equation}
(I_l^{(1)})_{\mu\mu}=ie{\bf E}\cdot \nabla_{\bf p} (\hat {\rm
G}_{0{\bf p}}^<)_{\mu\mu}+ R_{\mu \mu}, \label{EQ11}
\end{equation}
with $\mu=1,2$.

\section{Vanishing spin-Hall current}

We are interested in a spin current polarized along the
$z$-direction and flowing along the $y$ axis when a dc electric
field ${\bf E}$ is applied along the $x$ axis-{\it i.e.} $J_y^z$.
In spin basis, it relates to the single-particle spin current
operator $\check j_y^z$ by
\begin{equation}
J_y^z=\sum_{\bf p}\int \frac{d\omega}{2\pi} {\rm Tr}\left [\check
j_y^z{\rm Im}(\check {\rm G}_{1{\bf p}}^<)\right ].\label{YZ}
\end{equation}
There has been some debate about the definition of the spin
current $\check j_y^z$ in recent
years.\cite{Jin,Murakami,Yang,ZhangP} It seems reasonable to
define the spin current from a continuity equation relating to the
spin, as shown in Ref.\,\onlinecite{ZhangP}. Conventionally, the
single-particle spin current in spin basis is defined as $\check
j_y^z=\frac{1}{4}\{v_y,\sigma_z\}$ with the velocity operator
$v_y$. Zhang {\it et al.} suggested that the spin current operator
has an additional term arising from the spin torque density,
$(\check j_y^z)_{\rm add}=(y/2i)[\sigma_z,\check h_0 ({\bf
p})]$.\cite{ZhangP} Actually, the contribution from this term to
spin-Hall current vanishes in a homogeneous 2D Rashba electron
system to linear order in the dc field. This can be seen from the
fact that the additional spin-Hall current operator, explicitly
given by $(\check j_y^z)_{\rm add}=\alpha y {\bf p}\cdot {\bf
\sigma}=-i\alpha \frac{\partial}{\partial p_y} {\bf p}\cdot {\bf
\sigma}$, is invariant under inversion of momentum ${\bf p}$, {\it
i.e.} ${\bf p}\rightarrow -{\bf p}$. However, ${\bf p}$ reversal
leads to a sign change of the linear electric field correction to
the lesser Green's function: $\hat {\rm G}_{1,{-\bf p}}^<=-\hat
{\rm G}_{1{\bf p}}^<$ [this result can be seen from
Eq.\,(\ref{KE}) by inverting the momentum]. Based on this, it
follows from the ${\bf p}$ integration of Eq.\,(\ref{YZ}) that the
contribution to spin-Hall current arising from the additional
spin-current term vanishes. Hence, we only need to consider the
conventional component of the spin-Hall current operator.

Consequently, we write the spin current operator $J_y^z$ in the
simple helicity-basis description as\cite{Liu1}
\begin{equation}
J_y^z=\sum_{\bf p}\int \frac{d\omega}{2\pi}\frac{p_y}{m}{\rm
Im}(\hat{\rm G}^<_{\bf p})_{12}.\label{JYZZ}
\end{equation}
Substituting Eq.\,(\ref{EQ12}) into this equation, we obtain
\begin{widetext}
\begin{equation}
J_y^z= \sum_{\bf p}\int \frac{d\omega}{2\pi} \frac{1}{4
m\alpha}\sin\phi_{\bf p}e{\bf E}\cdot \nabla_{\bf p} \phi_{\bf
p}{\rm Im}\left [(\hat {\rm G}_{0{\bf p}}^<)_{11}-(\hat {\rm
G}_{0{\bf p}}^<)_{22}\right ] +\sum_{\bf p}\int
\frac{d\omega}{2\pi}\frac 1{2m\alpha}\sin \phi_{\bf p} {\rm Im}
(I^{(1)}_l)_{12}. \label{JYZ}
\end{equation}
In this, the $\omega$ integration over the term $R$ vanishes.

To further simplify the expression for $J_y^z$, we analyze the
second term on the right-hand side of Eq.\,(\ref{JYZ}). In
connection with this, we find that there is a sum relation among
the elements of $I^{(1)}_l$ in the self-consistent Born
approximation:
\begin{equation}
\sum_{\bf p}\int \frac{d\omega}{2\pi}\sin \phi_{\bf p} {\rm Im}
(I^{(1)}_l)_{12}= \sum_{\bf p}\int \frac{d\omega}{2\pi}\frac{\cos
\phi_{\bf p}}{2} \left [{\rm Re} (I^{(1)}_l)_{11}-{\rm Re}
(I^{(1)}_l)_{22}\right ].\label{SR}
\end{equation}
Furthermore, separate relations also hold for each component of
$I^{(1)}_l$:
\begin{equation}
\sum_{\bf p}\int\frac{d\omega}{2\pi}\sin \phi_{\bf p} {\rm Im}
(I^{(1)}_{l1})_{12}= \sum_{\bf p}\int
\frac{d\omega}{2\pi}\frac{\cos \phi_{\bf p}}{2} \left [{\rm Re}
(I^{(1)}_{l1})_{11}-{\rm Re} (I^{(1)}_{l1})_{22}\right
],\label{SR1}
\end{equation}
\begin{equation}
\sum_{\bf p}\int\frac{d\omega}{2\pi}\sin \phi_{\bf p} {\rm Im}
(I^{(1)}_{l3})_{12}= \sum_{\bf p}\int
\frac{d\omega}{2\pi}\frac{\cos \phi_{\bf p}}{2} \left [{\rm Re}
(I^{(1)}_{l2})_{11}-{\rm Re} (I^{(1)}_{l2})_{22}\right
].\label{SR2}
\end{equation}
The detailed proof of these equations is presented in Appendix D.

Based on the sum relation Eqs.\,(\ref{SR}) and (\ref{EQ11})
for diagonal components of $I_l^{(1)}$, we obtain
\begin{eqnarray}
J_y^z&=&\sum_{\bf p}\int \frac{d\omega}{2\pi} \frac{\sin\phi_{\bf
p}}{4m\alpha}e{\bf E}\cdot \nabla_{\bf p} \phi_{\bf p}{\rm
Im}\left [(\hat {\rm G}_{0{\bf p}}^<)_{11}-(\hat {\rm G}_{0{\bf
p}}^<)_{22}\right ]\nonumber\\
 &&-\sum_{\bf p}\int \frac{d\omega}{2\pi}
\frac{\cos\phi_{\bf p}}{4m\alpha}e{\bf E}\cdot \left (\nabla_{\bf
p} {\rm Im}\left [(\hat {\rm G}_{0{\bf p}}^<)_{11}-(\hat {\rm
G}_{0{\bf p}}^<)_{22}\right ]\right) \nonumber\\
&=&-\sum_{\bf
p}\int \frac{d\omega}{2\pi} \frac{e{\bf E}\cdot \nabla_{\bf p}
\cos\phi_{\bf p}}{4m\alpha} {\rm Im}\left [(\hat {\rm G}_{0{\bf
p}}^<)_{11}-(\hat {\rm G}_{0{\bf p}}^<)_{22}\right ]
\nonumber\\
 &&-\sum_{\bf
p}\int \frac{d\omega}{2\pi} \frac{\cos\phi_{\bf p}}{4m\alpha}e{\bf
E}\cdot \left (\nabla_{\bf p} {\rm Im}\left [(\hat {\rm G}_{0{\bf
p}}^<)_{11}-(\hat {\rm G}_{0{\bf p}}^<)_{22}\right ]\right)
.\label{ZERO}
\end{eqnarray}
\end{widetext}
It is apparent from Eq.\,(\ref{ZERO}) that the integrand on its
right-hand side becomes a total derivative with respect to ${\bf
p}$. Consequently, the vanishing of the spin-Hall current is
obtained immediately upon the momentum integration. We note that
this analytical result is significantly different from that of the
numerical analysis of Sugimato {\it et al.},\cite{Nagaosa} which
purports to exhibit a nonvanishing spin-Hall current. Our analysis
shows that this is not the case even with the "new" spin current
term included.

It is obvious from Eq.\,(\ref{ZERO}) that the spin-Hall current
vanishes independently of the strength of SO coupling, of the
specific form of the isotropic scattering potential, and of the
lattice temperature. Moreover, it is valid even for arbitrary
broadening of the DOS. However, this result holds only in the
diffusive regimes-namely, in the regime $l_Dk_{\rm F} > 1$ ($l_D$
is diffusion length and $k_{\rm F}$ is the Fermi wave vector),
because we restricted treatment of the electron-impurity
scattering to the self-consistent Born approximation (otherwise,
contributions from the maximally crossed diagrams of
electron-impurity scattering would have to be taken into account).

\section{Conclusions}
The spin-Hall effect in 2D diffusive Rashba systems has been
investigated by means of a quantum Boltzmann approach in this
paper. In the self-consistent Born approximation, we have
considered all effects induced by electron-impurity
scattering-namely, quasiclassical relaxation, collisional
broadening, and the intracollisional field effect. We have proven
analytically that the spin-Hall current vanishes, irrespective of
the strength of SO coupling, of the broadening of DOS, and of the
form of the isotropic scattering potential. The sum relation of
the scattering terms in the helicity basis has been analyzed.

Our treatment is applicable at nonvanishing temperatures (not
restricted to $T=0$), and it is not restricted in anyway to
short-range scattering potentials. We have also made it clear that
the spin-Hall effect vanishes even when correctly considering the influence
of the electric field in the scattering process
(intracollisional field effect).

It should be noted that this proof of vanishing of spin-Hall current also
applies for a Dresselhaus linear-momentum  SO coupling, since it
relates simply to the Rashba SO interaction by a global
transformation. Furthermore, we have also examined the spin-Hall
effect when both the Rashba and Dresselhaus SO interactions are
active, again finding zero spin-Hall current.

Nevertheless, we note that our proof of vanishing of the spin-Hall
effect has its validity limited to 2D systems with linear-momentum
SO coupling. Also, this proof requires $\check h_0$ to be
parabolic in the absence of SO coupling. Otherwise, the spin-Hall
effect should be examined anew. This is to say that our result
does not conflict with the nonzero experimental spin-Hall
observations\cite{Kato, Wunderlich}: the Hamiltonians of bulk and
2D heavy-hole systems in which nonvanishing spin-Hall currents
were observed differ significantly from the Hamiltonian $\check
h_0 ({\bf p})$ studied in this paper.

{\it Note added}. After this work was completed and submitted, an
unpublished paper of Krotkov and Das Sarma reported a nonvanishing
spin-Hall effect in 2D Rashba electron systems by considering a
nonparabolic effect.\cite{Sarma}

\begin{acknowledgments}
One of the authors (SYL) would like to gratefully acknowledge the
invaluable discussions with Dr. M. W. Wu, Dr. W. Xu, Dr. W. S.
Liu, and Dr. Y. Chen. This work was supported by projects of the
National Science Foundation of China and the Shanghai Municipal
Commission of Science and Technology and by the Youth Scientific
Research Startup Founds of SJTU. N.J.M.H. is supported by the
Department of Defense through the DURINT program administered by
the U.S. Army Research Office, DAAD Grant No. 19-01-1-0592.
\end{acknowledgments}

\appendix
\section{Nonequilibrium Green's functions in spin basis}

In spin space, the $2\times 2$ nonequilibrium Green's functions
${\check {\rm G}}({\bf r}_1,\tau_1;{\bf r}_2,\tau_2)$ obey Dyson
equations given by\cite{Jauho}
\begin{widetext}
\begin{eqnarray}
\left \{i\frac{\partial}{\partial \tau_1}+\frac{1}{2m}{\vec
\nabla}^2_{{\bf r}_1} +i\alpha {\nabla}_{{\bf r}_1}\cdot ({\bf
n}\times {\bf \sigma}) +e{\bf E}\cdot {\bf r}_1 \right
\}\check{\rm G}({\bf r}_1,\tau_1;{\bf r}_2,\tau_2)=\nonumber
\end{eqnarray}
\begin{equation}
\delta(1-2)+\int_Cd\tau '\int d {\bf r}' \check \Sigma({\bf r}_1,\tau_1;{\bf r}',\tau ')\check {\rm G}({\bf r}',\tau ';{\bf r}_2,\tau_2),
\label{Eq3}
\end{equation}
\begin{eqnarray}
\check{\rm G}({\bf r}_1,\tau_1;{\bf r}_2,\tau_2) \left
\{-i\frac{\partial}{\partial \tau_2}+\frac{1}{2m}{\vec
\nabla}^2_{{\bf r}_2} -i\alpha {\nabla}_{{\bf r}_2}\cdot ({\bf
n}\times {\bf \sigma}) +e{\bf E}\cdot {\bf r}_2 \right
\}=\nonumber
\end{eqnarray}
\begin{equation}
\delta(1-2)+\int_Cd\tau '\int d {\bf r}' \check {\rm G}({\bf r}_1,\tau_1;{\bf r}',\tau ')\check \Sigma
({\bf r}',\tau ';{\bf r}_2,\tau_2).\label{Eq4}
\end{equation}
\end{widetext}
In these equations, the electron-impurity interaction is embedded
in the self-energies, $\check \Sigma ({\bf r}',\tau ';{\bf
r}_2,\tau_2)$.

To further simplify Eqs.\,(\ref{Eq3}) and (\ref{Eq4}), we
introduce relative and center-of-mass variables ${\bf r}={\bf
r}_1-{\bf r}_2$, $\tau=\tau_1-\tau_2$, ${\bf R}=({\bf r}_1+{\bf
r}_2)/2$, and $T=(\tau_1+\tau_2)/2$. Following this, we construct
gauge-invariant retarded and lesser Green's functions $\check {\rm
G}^{r,<}_{\bf p}$ in momentum-frequency space following
established procedures.\cite{Jauho} Under homogeneous and
steady-state conditions, these Green's functions satisfy the
following equations:
\begin{widetext}
\begin{equation}
2\left [\omega-\frac {p^2}{2m}\right ]\check {\rm G}^r_{\bf p}
+i\alpha\left [ i{\bf p}\cdot \check{\bf B}^r+\frac 12 e{\bf
E}\cdot\frac{\partial}{\partial \omega} \check{\bf A}^r_{\bf
p}\right ]=2+\check I_{r1} +\check I_{r2},\label{DS1}
\end{equation}
\begin{equation}
i\left [e{\bf E}\cdot \left ({\nabla}_{\bf p}+{\bf p}\frac
\partial {\partial \omega}\right )\right ] \check{\rm G}^<_{\bf p}
+i\alpha\left [ i{\bf p}\cdot \check{\bf A}^<+\frac 12 e{\bf
E}\cdot\frac{\partial}{\partial \omega} \check{\bf B}_{\bf
p}^<\right ]=\check I_{l1} +\check I_{l2},\label{DS2}
\end{equation}
\end{widetext}
where $\check{\bf A}^{r,<}_{\bf p}\equiv [{\bf n}\times {\bf
\sigma}, {\check {\rm G}}^{r,<}_{\bf p}]$ and $\check{\bf
B}^{r,<}_{\bf p}\equiv \{{\bf n}\times {\bf \sigma}, {\check {\rm
G}}^{r,<}_{\bf p}\}$ arise from the Rashba spin-orbit interaction
term in the Hamiltonian. The first terms $\check I_{r1}$ and
$\check I_{l1}$ on the right-hand sides of Eqs.\,(\ref{DS1}) and
(\ref{DS2}), respectively, do not involve  the explicit appearance
of the electric field
\begin{equation}
\check I_{r1}=\check{\Sigma}^r_{\bf p}\check{\rm G}^r_{\bf p} +
\check{\rm G}^r_{\bf p} \check{\Sigma}^r_{\bf p},
\end{equation}
\begin{equation}
\check I_{l1}= \check{\Sigma}^r_{\bf p}\check{\rm G}^<_{\bf p} -
\check{\rm G}^<_{\bf p} \check{\Sigma}^a_{\bf p} - \check{\rm
G}^r_{\bf p} \check{\Sigma}^<_{\bf p} +
   \check{\Sigma}^<_{\bf p} \check{\rm G}^a_{\bf p}.
\end{equation}
The terms $\check I_{r2}$ and $\check I_{l2}$ arise from the
first-order gradient expansion involving explicit linear
dependence on the electric field,
\begin{eqnarray}
\check I_{r2}&=&\frac{i}{2}e{\bf E}\cdot \left ( \nabla_{\bf
p}\check{\Sigma}^r_{\bf p} \frac{\partial}{\partial
\omega}\check{\rm G}^r_{\bf p}-\frac{\partial}{\partial \omega}
\check{\Sigma}^r_{\bf p}
\nabla_{\bf p}\check{\rm G}^r_{\bf p}\right .\nonumber\\
&&\left .+\nabla_{\bf p}\check{\rm G}^r_{\bf p}
\frac{\partial}{\partial \omega}\check{\Sigma}^r_{\bf
p}-\frac{\partial}{\partial \omega} \check{\rm G}^r_{\bf p}
\nabla_{\bf p}\check{\Sigma}^r_{\bf p}
 \right ),
\end{eqnarray}
\begin{eqnarray}
\check I_{l2}=\frac i2 e{\bf E}\cdot \left(\nabla_{\bf
p}\check{\Sigma}^r_{\bf p} \frac{\partial}{\partial
\omega}\check{\rm G}^<_{\bf p}-\frac{\partial}{\partial \omega}
\check{\Sigma}^r_{\bf p} \nabla_{\bf p}\check{\rm G}^<_{\bf p}
\right .\nonumber\\
\left . - \nabla_{\bf p}\check{\rm G}^<_{\bf
p}\frac{\partial}{\partial \omega}\check{\Sigma}^a_{\bf p}
+\frac{\partial}{\partial \omega} \check{\rm G}^<_{\bf p}
\nabla_{\bf p}\check{\Sigma}^a_{\bf p} - \nabla_{\bf p}\check{\rm
G}^r_{\bf p} \frac{\partial}{\partial \omega}\check{\Sigma}^<_{\bf
p}
\right .\nonumber\\
\left . +\frac{\partial}{\partial \omega} \check{\rm G}^r_{\bf p}
\nabla_{\bf p}\check{\Sigma}^<_{\bf p}+\nabla_{\bf
p}\check{\Sigma}^<_{\bf p} \frac{\partial}{\partial
\omega}\check{\rm G}^a_{\bf p}-\frac{\partial}{\partial \omega}
\check{\Sigma}^<_{\bf p} \nabla_{\bf p}\check{\rm G}^a_{\bf p}
\right ).
\end{eqnarray}

Incidentally, ignoring the collisional broadening, i.e., the terms
involving $\partial/\partial \omega$, and employing the
generalized Kadanoff-Baym ansatz, Eq. (\ref{DS1}) reduces to the
kinetic equations of the distribution functions presented in the
previous studies.\cite{Add}

\section{Retarded Green's function}

To analyze the properties of the retarded Green's function, we begin from
the Dyson equation in the helicity basis:
\begin{widetext}
\begin{equation}
2\left [\omega-\frac {p^2}{2m}\right ]\hat {\rm G}^r_{\bf p}
+\alpha p \{\sigma_z, \hat {\rm G}^r_{\bf p}\}+\frac {i\alpha} 2 e{\bf E}\cdot \frac{\partial}{\partial \omega}\hat{\bf A}^r_{\bf p}=2+\hat I_{r1}
+\hat I_{r2}+\hat I_{r3}.
\end{equation}
\end{widetext}
This equation is obtained from Eq.\,(\ref{DS1}) by the
transformation from spin basis to the helicity basis.
The quantities $\hat I_{r1}$ and $\hat I_{r2}$
are the helicity-basis analogues of the
quantities $\check I_{r1}$ and $\check I_{r2}$, respectively.
$\hat I_{r3}$ has a form similar to  $\hat I_{r2}$, but
with $\frac i 2\nabla_{\bf p}\phi_{\bf p}[\cdots,\sigma_x]$ in place of the operator $\nabla_{\bf p}$.

Obviously, in the helicity basis, the noninteracting retarded
Green's function $\hat g^r_{\bf p}$ is diagonal
\begin{equation}
\hat g^r_{\bf p}={\rm diag}\left
((\omega-\varepsilon_1(p)+i\delta)^{-1},
(\omega-\varepsilon_2(p)+i\delta)^{-1}\right ).
\end{equation}
Including collisional broadening in the definition of the
unperturbed retarded Green's function $\hat{\rm G}^r_{0{\bf p}}$,
it obeys the Dyson equation in the absence of the electric field,
as
\begin{equation}
\left [ \omega-\frac {p^2}{2m}\right ]\hat{\rm G}^r_{0{\bf p}}+\frac{\alpha
p}2\{\sigma_z,\hat {\rm G}^r_{0{\bf p}}\} =1+\frac 12(\hat \Sigma^r_{0{\bf p}} \hat
{\rm G}^r_{0{\bf p}}+ \hat {\rm G}^r_{0{\bf p}}\hat \Sigma^r_{0{\bf p}}).\label{DER0}
\end{equation}
The solution of this equation is also a diagonal matrix, $(\hat
{\rm G}^r_{0{\bf p}})_{12}=(\hat {\rm G}^r_{0{\bf p}})_{21}=0$, with elements
independent of the direction of momentum.
To verify this, we examine the form of the
self-energy $\hat \Sigma^r_{0{\bf p}}$ below.

Let us first discuss the forms of self-energies
$\hat \Sigma^{r,<}_{{\bf p}}$ in general.
Installing the actual form of matrix ${U}_{\bf p}$ into Eq.\,(\ref{SE}), we
find
\begin{eqnarray}
{\hat \Sigma}^{r,<}_{\bf p}&=&\frac 12 n_i\sum_{{\bf k}}|V({\bf
p}-{\bf k})|^2
\left \{ a_1 \hat {\rm G}^{r,<}_{\bf k}\right .\nonumber \\
&&\left .+a_2{\hat \sigma}_x\hat {\rm G}^{r,<}_{\bf k}{\hat
\sigma}_x+ia_3[{\hat \sigma}_x,\hat {\rm G}^{r,<}_{\bf k}]\right
\}.
\end{eqnarray}
Explicitly, the self-energies can also be rewritten as
\begin{eqnarray}
(\hat \Sigma^{r,<}_{\bf p})_{\mu\nu}= \frac {n_i} 2 \sum_{\bf
k}|V({\bf p}-{\bf k})|^2\left \{ a_1(\hat {\rm G}^{r,<}_{\bf
k})_{\mu\nu}+
\right .\nonumber \\
\left .
a_2(\hat {\rm G}^{r,<}_{\bf k})_{\bar \mu \bar\nu}
 +ia_3[(\hat {\rm G}^{r,<}_{\bf k})_{\bar \mu\nu}
-(\hat {\rm G}^{r,<}_{\bf k})_{\mu\bar\nu}]\right \}.\label{SER}
\end{eqnarray}
In these expressions, $\bar \mu=3-\mu$ and $a_i$($i=1,2,3$) are
factors associated with the directions of momenta, $a_1=1+\cos
(\phi_{\bf p}-\phi_{\bf k})$, $a_2=1-\cos (\phi_{\bf p}-\phi_{\bf
k})$, and $a_3=\sin (\phi_{\bf p}-\phi_{\bf k})$.

From Eq.\,(\ref{SER}), we see that the diagonal elements of the
Green's functions appearing in nondiagonal elements of ${\hat
\Sigma}^{r,<}_{\bf p}$ are always combined with the factor $a_3$ .
If the solution of Eq.\,(\ref{DER0}), $\hat {\rm G}^{r}_{0{\bf
p}}$, is diagonal and independent of the direction of momentum,
the self-energy term involving $a_3$ vanishes under momentum
integration. Hence, the self-energy, ${\hat \Sigma}^{r}_{0{\bf
p}}$, also becomes diagonal and independent of momentum direction.
As a result, Eq.\,(\ref{DER0}) reduces to a
momentum-direction-independent equation for the diagonal elements
of $\hat{\rm G}^{r}_{0{\bf p}}$. This implies that the assumed
form of $\hat{\rm G}^{r}_{0{\bf p}}$ is consistent with
Eq.\,(\ref{DER0}). From uniqueness of the solution of the Dyson
equation we immediately conclude that such a $\hat{\rm
G}^{r}_{0{\bf p}}$ is just the needed solution. It should be noted
that, physically, the momentum-direction independence of the
unperturbed retarded Green's function is associated with the
rotational symmetry of the unperturbed Hamiltonian in helicity
basis.

Considering the angular independence of $\hat {\rm G}^r_{0{\bf p}}$,
an important relation can be derived:
\begin{equation}
\sum_{\bf k}|V({\bf p}-{\bf k})|^2\sin(\phi_{\bf p}-\phi_{\bf
k})[(\hat {\rm G}^r_{0{\bf k}})_{22} -(\hat {\rm G}^r_{0{\bf
k}})_{11}]=0.\label{TT}
\end{equation}
This equation is needed to prove the sum relation of the scattering term (Appendix D).

To the first order in the dc field, the Dyson equation for the
linear electric field correction to the retarded Green's function,
$\hat {\rm G}^r_{1{\bf p}}=\hat {\rm G}^r_{\bf p}-\hat {\rm G}^r_{0{\bf p}}$, can be written as
\begin{widetext}
\begin{eqnarray}
2\left [\omega-\frac{p^2}{2m}+(-1)^\mu\delta_{\mu\nu}\alpha p\right ](\hat{\rm G}^r_1)_{\mu\nu}
+(1-\delta_{\mu\nu})\frac {\alpha eE}{2}\sin\phi_{\bf p}\frac{\partial}{\partial\omega}
\left [(\hat{\rm G}^r_0)_{11}-(\hat{\rm G}^r_0)_{22}\right ]\nonumber\\
=(\hat I_{r1}^{(1)})_{\mu\nu}+(\hat I_{r2}^{(1)})_{\mu\nu}
+(\hat I_{r3}^{(1)})_{\mu\nu}.\label{DER1}
\end{eqnarray}
$\hat I_{r1}^{(1)}$, $\hat I_{r2}^{(1)}$, and $\hat I_{r3}^{(1)}$
are the linear electric field parts of $\hat I_{r1}$, $\hat
I_{r2}$ and $\hat I_{r3}$. $\hat I_{r1}^{(1)}$ has the form
\begin{equation}
\hat I_{r1}^{(1)}=2
\left (
\begin{array}{cc}
(\hat{\rm
G}^r_{0{\bf p}})_{11}(\hat{\Sigma}^r_{1{\bf p}})_{11}& [(\hat{\rm G}^r_{0{\bf p}})_{11}+(\hat{\rm
G}^r_{0{\bf p}})_{22}](\hat{\Sigma}^r_{1{\bf p}})_{12}
\\

[(\hat{\rm G}^r_{0{\bf p}})_{11}+(\hat{\rm
G}^r_{0{\bf p}})_{22}](\hat{\Sigma}^r_{1{\bf p}})_{21} &
(\hat{\rm
G}^r_{0{\bf p}})_{22}(\hat{\Sigma}^r_{1{\bf p}})_{22}
\end{array}
\right )+(\hat{\rm G}^r_{0{\bf p}}\rightarrow\hat{\Sigma}^r_{0{\bf p}},\hat{\Sigma}^r_{1{\bf p}}\rightarrow\hat{\rm G}^r_{1{\bf p}} ),
\end{equation}
where $\hat{\Sigma}^{r}_{1\bf p}=n_i\sum_{{\bf k}}\hat T({\bf
p},{\bf k})\hat{\rm G}^r_{1\bf k}\hat T^+({\bf p},{\bf k})$. $\hat
I_{r2}^{(1)}$ vanishes due to the diagonal nature of the
unperturbed retarded Green's function and self-energy. $\hat
I_{r3}^{(1)}$ is off-diagonal and symmetric, $(\hat
I_{r3}^{(1)})_{12}=(\hat I_{r3}^{(1)})_{21}$. From this Dyson
equation and the explicit forms of self-energies (\ref{SER}), we
find that $\hat {\rm G}_{1{\bf p}}^r$ becomes an off-diagonal
matrix with symmetric elements, $(\hat {\rm G}_{1{\bf p}}^r)_{12}
=(\hat {\rm G}_{1{\bf p}}^r)_{21}$. Also, $\hat {\Sigma}_{1{\bf
p}}^r$ and hence $\hat I_{r1}^{(1)}$ have vanishing diagonal
elements. At the same time, Eq.\,(\ref{DER1}) reduces to an equation for nondiagonal
elements of $\hat {\rm G}_{1{\bf p}}^r$.

Actually, $\hat {\rm G}_{1{\bf p}}^r$ depends on the
direction of momentum through a sine function. The expression for $\hat {\rm G}_{1{\bf p}}^r$ can be formally written as
\begin{eqnarray}
(\hat {\rm G}_{1{\bf p}}^r)_{\mu\bar\mu}
&=&\frac{eE\sin \phi_{\bf p}}{\omega-p^2/2m-\tau_{p}^{-1}}\nonumber\\
&&\times \left \{-\frac{\alpha}{2}\frac{\partial}
{\partial \omega}[(\hat{\rm G}^r_{0{\bf p}})_{11}-(\hat{\rm G}^r_{0{\bf p}})_{22}]
+[(\hat{\rm G}^r_{0{\bf p}})_{11}+(\hat{\rm G}^r_{0{\bf p}})_{22}]\Lambda_{ p}\right \},\label{DER2}
\end{eqnarray}
with $\Lambda_{p}=2n_i\sum_{\bf k}|V({\bf p}-{\bf
k})|^2\cos(\phi_{\bf p}-\phi_{\bf k})(\hat {\cal G}_{1{
p}}^r)_{12}$, $\tau^{-1}_{p}=2n_i\sum_{\bf k}|V({\bf p}-{\bf
k})|^2[(\hat{\rm G}^r_{0{\bf p}})_{11}+(\hat{\rm G}^r_{0{\bf
p}})_{22}]$, and $\hat {\cal G}_{1 p}^r=\hat {\rm G}_{1{\bf
p}}^r/[eE\sin \phi_{\bf p}]$ as a quantity independent of the
direction of momentum. In the case of short-range disorders, the
$\Lambda_{p}$ and $\tau_{p}$ become momentum-independent
constants. From Eq.\,(\ref{DER2}) it is obvious that $\hat {\rm
G}_{1{\bf p}}^r$ does not vanish.

We note that the electron-impurity scattering
is not essential for the nonvanishing of $\hat {\rm G}_{1{\bf p}}^r$.
In this, the second term on the left-hand side of Eq.\,(\ref{DER1})
plays a key role. In the absence of the electron-impurity
collisions, $\hat {\rm G}_{1{\bf p}}^r$ can be obtained analytically as
\begin{equation}
(\hat{\rm G}^r_{1{\bf p}})_{12}=(\hat{\rm G}^r_{1{\bf p}})_{21}=-
\frac{\alpha eE}{4(\omega-{p^2}/{2m})}
\sin\phi_{\bf p}\frac{\partial}{\partial\omega}
\left [(\hat{\rm G}^r_{0{\bf p}})_{11}-(\hat{\rm G}^r_{0{\bf p}})_{22}\right ].
\label{DDG}
\end{equation}
Incidentally, from Eq.\,(\ref{DDG}) we can derive
the retarded Green's function presented in our previous work.\cite{Liu1}
Using the equality $(\hat{\rm G}^r_{0{\bf p}})\equiv (\hat{\rm g}^r_{0{\bf p}})$
in the absence of electron-impurity collisions, we rewrite Eq.\,(\ref{DDG}) as
$(\hat{\rm G}^r_{1{\bf p}})_{12}=(\hat{\rm G}^r_{1{\bf p}})_{21}=
{\cal A}_1+{\cal A}_2$ with
\begin{equation}
{\cal A}_1=
-\frac{eE}{4\alpha p^2}\sin\phi_{\bf p}
\left [(\hat{\rm G}^r_{0{\bf p}})_{11}-(\hat{\rm G}^r_{0{\bf p}})_{22}\right ]
\end{equation}
and
\begin{equation}
{\cal A}_2=-\frac{eE}{4 p}\sin\phi_{\bf p}\frac{\partial}{\partial\omega}
\left [(\hat{\rm G}^r_{0{\bf p}})_{11}+(\hat{\rm G}^r_{0{\bf p}})_{22}\right ].
\end{equation}
${\cal A}_1$ is just the term that we employed in Ref.\,\onlinecite{Liu1},
and ${\cal A}_2$ is a total derivative with respect to $\omega$.
In the absence of electron-impurity scattering, the term ${\cal A}_2$ has no effect on
the spin dynamics of the considered system, because the quantity directly associated with
spin-Hall current, $\hat {\rm G}_{1{\bf p}}^<$, takes the form
\begin{equation}
\hat{\rm G}_{1{\bf p}}^<=-2in_{\rm F}(\omega){\rm Im} {\cal A}_1+{\cal R},
\end{equation}
with ${\cal R}$ as a remainder term which can be expressed as a
total derivative with respect to $\omega$ [this result is obtained
from Eq.\,(\ref{EQ12})].

\section{explicit expression for $\hat I_l^{(1)}$}

The scattering term $\hat I_l^{(1)}$ can be expressed as the sums
of $\hat I_{l1}^{(1)}$, $\hat I_{l2}^{(1)}$, and $\hat
I_{l3}^{(1)}$, the corresponding linear electric field parts of
$\hat I_{l1}$, $\hat I_{l2}$, and $\hat I_{l3}$. Further, $\hat
I_{l1}^{(1)}$ can be rewritten as $\hat I_{l1}^{(1)}=\hat F_1+\hat
F_2$, with an off-diagonal matrix $\hat F_1 \equiv \hat
\Sigma_{1{\bf p}} ^r \hat{\rm G}_{0{\bf p}}^< -\hat{\rm G}_{0{\bf
p}}^< \hat \Sigma_{1{\bf p}} ^a -\hat{\rm G}_{1{\bf p}}^r\hat
\Sigma_{0{\bf p}} ^<+\hat \Sigma_{0{\bf p}} ^< \hat{\rm G}_{1{\bf
p}}^a $, or explicitly,
\begin{equation}
(\hat F_{1})_{\mu\bar\mu}=(\hat \Sigma_{1{\bf p}}^r)_{\mu\bar\mu}(\hat
{\rm G}_{0{\bf p}}^<)_{\bar\mu\bar\mu}-(\hat \Sigma_{1{\bf p}}^a)_{\mu\bar\mu}
(\hat {\rm G}_{0{\bf p}}^<)_{\mu\mu}
-(\hat {\rm G}_{1{\bf p}}^r)_{\mu\bar\mu}(\hat
{\Sigma}_{0{\bf p}}^<)_{\bar\mu\bar\mu}+(\hat {\rm
G}^a_{1{\bf p}})_{\mu\bar\mu}(\hat {\Sigma}_{0{\bf p}}^<)_{\mu\mu},
\end{equation}
and a general matrix $\hat F_2$ with elements
\begin{equation}
(\hat F_2)_{\mu\mu}=2i[{\rm Im} ({\hat \Sigma}_{0{\bf p}}^r)_{\mu\mu}
(\hat{\rm G}_{1{\bf p}}^<)_{\mu\mu}- {\rm Im} ({\hat {\rm G}}_{0{\bf p}}^r)_{\mu\mu}
(\hat{\Sigma}_{1{\bf p}}^<)_{\mu\mu}],
\end{equation}
\begin{equation}
(\hat F_2)_{\mu\bar\mu}=[({\hat \Sigma}_{0{\bf p}}^r)_{\mu\mu}-({\hat
\Sigma}_{0{\bf p}}^a)_{\bar\mu\bar\mu}] (\hat{\rm G}_{1{\bf p}}^<)_{\mu\bar\mu}-
[({\hat {\rm G}}_{0{\bf p}}^r)_{\mu\mu}-({\hat {\rm
G}}_{0{\bf p}}^a)_{\bar\mu\bar\mu}] (\hat{\Sigma}_{1{\bf p}}^<)_{\mu\bar\mu},
\end{equation}
where linear electric field correction to the lesser self-energy,
$\hat{\Sigma}^{<}_{1\bf p}$, takes the form:
$\hat{\Sigma}^{<}_{1\bf p}=n_i\sum_{{\bf k}}\hat T({\bf p},{\bf
k})\hat{\rm G}^<_{1\bf k}\hat T^+({\bf p},{\bf k})$. Furthermore,
the diagonal $\hat I_{l2}^{(1)}$ and off-diagonal $I_{l3}^{(1)}$
matrices have the forms
\begin{eqnarray}
(\hat I_{l2}^{(1)})_{\mu\mu}&=&ie{\bf E}\cdot \left (\nabla_{\bf
p}{\rm Re} (\hat {\Sigma}^r_{0{\bf p}})_{\mu\mu}
 \frac{\partial}{\partial \omega}(\hat {\rm G}^<_{0{\bf p}})_{\mu\mu}-
\frac{\partial}{\partial \omega} {\rm Re}(\hat{\Sigma}^r_{0{\bf p}})_{\mu\mu}
\nabla_{\bf p}(\hat{\rm G}^<_{0{\bf p}})_{\mu\mu}
\right .
\nonumber\\
&& \left . - \nabla_{\bf p}{\rm Re}(\hat{\rm G}^r_{0{\bf p}})_{\mu\mu}
\frac{\partial}{\partial \omega}(\hat{\Sigma}^<_{0{\bf p}})_{\mu \mu}
+\frac{\partial}{\partial \omega} {\rm Re}(\hat{\rm G}^r_{0{\bf p}})_{\mu
\mu} \nabla_{\bf p}(\hat {\Sigma}^<_{0{\bf p}})_{\mu\mu} \right
),\label{IL2}
\end{eqnarray}
\begin{eqnarray}
(\hat I_{l3}^{(1)})_{\mu\bar\mu}&=&(-1)^{\mu}\frac {e}{4}{\bf
E}\cdot \nabla_{\bf p} \phi_{\bf p} \sum_{\nu}(-1)^{\nu}\left [
(\hat {\rm G}^<_{0{\bf p}})_{\nu\nu} \frac{\partial}{\partial \omega}{\rm
Re}(\hat {\Sigma}^r_{0{\bf p}})_{\nu\nu}-
 \frac{\partial}{\partial \omega}(\hat {\rm G}^<_{0{\bf p}})_{\nu\nu}
{\rm Re}(\hat {\Sigma}^r_{0{\bf p}})_{\nu\nu}\right .
\nonumber\\
&& \left . -\frac{\partial}{\partial\omega}{\rm Re}(\hat {\rm
G}^r_{0{\bf p}})_{\nu\nu} (\hat {\Sigma}^<_{0{\bf p}})_{\nu\nu}+ {\rm Re}(\hat {\rm
G}^r_{0{\bf p}})_{\nu\nu}
 \frac{\partial}{\partial \omega}
(\hat {\Sigma}^<_{0{\bf p}})_{\nu\nu}\right ]+(\hat
R_2)_{\mu\bar\mu},\label{IL3}
\end{eqnarray}

From the definition of the lesser Green's function, we know that
$(\hat {\rm G}^{<}_{1{\bf p}})_{\mu\nu}=-(\hat {\rm G}^{<}_{1{\bf
p}})^*_{\nu\mu}$. This implies that its diagonal elements are pure
imaginary-{\it i.e.}, ${\rm Re}(\hat {\rm G}^{<}_{1{\bf
p}})_{\mu\mu}=0$. Hence, we see from Eq.\,(\ref{SER}) that the
diagonal elements of the linear electric field part of the
self-energy, $\Sigma^<_{1{\bf p}}$, are also pure imaginary.
Consequently, the diagonal matrix $\hat F_2$ is pure imaginary, as
well as the matrix $\hat I^{(1)}_{l2}$.

\section{Proof of $\hat I_l^{(1)}$ sum relation}

First we prove Eq.\,(\ref{SR1}). Since ${\rm Im} ({\hat
I}_{l1}^{(2)})_{12}$ is the sum of $(\hat F_1)_{12}$ and ${\rm
Im}(\hat F_2)_{12}$, we first consider the integral ${\cal
I}_1\equiv \sum_{\bf p}\sin\phi_{\bf p}(\hat F_1)_{12}$.
Explicitly, it takes the form
\begin{equation}
{\cal I}_1=\sum_{\bf p}\sin \phi_{\bf p}\left \{ {\rm
Re}(\hat\Sigma_{1{\bf p}}^{r})_{12}[{\rm Im} (\hat{\rm G}_{0{\bf
p}}^<)_{22}-{\rm Im} (\hat{\rm G}_{0{\bf p}}^<)_{11}] -{\rm Re}
(\hat{\rm G}_{1{\bf p}}^r)_{12}[{\rm Im} (\hat\Sigma_{0{\bf
p}}^<)_{22}-{\rm Im} (\hat\Sigma_{0{\bf p}}^<)_{11}] \right
\}.\label{FF1}
\end{equation}
Substituting the self-energies into Eq.\,(\ref{FF1}), we have
\begin{eqnarray}
{\cal I}_1&=&\sum_{{\bf p}{\bf k}}\sin\phi_{\bf p}|V({\bf p}-{\bf k})|^2\left \{
{\rm Re}(\hat{\rm G}_{1{\bf k}}^{r})_{12}[{\rm Im} (\hat{\rm G}_{0{\bf p}}^<)_{22}-{\rm Im} (\hat{\rm G}_{0{\bf p}}^<)_{11}]\right.\nonumber\\
&&\left.
+\cos(\phi_{\bf p}-\phi_{\bf k}){\rm Re} (\hat{\rm G}_{1{\bf p}}^r)_{12}[{\rm Im} (\hat{\rm G}_{0{\bf k}}^<)_{11}-{\rm Im} (\hat{\rm G}_{0{\bf k}}^<)_{22}]
\right \}.
\end{eqnarray}
Interchanging the dummy integration variables ${\bf p}$ and ${\bf
k}$, ${\cal I}_1$ can be rewritten as
\begin{eqnarray}
{\cal I}_1&=&\sum_{{\bf p}{\bf k}}|V(|{\bf p}-{\bf k}|)|^2
(\sin\phi_{\bf p}-\sin\phi_{\bf k}\cos(\phi_{\bf p}-\phi_{\bf k}))
{\rm Re}(\hat{\rm G}_{1{\bf k}}^{r})_{12}[{\rm Im} (\hat{\rm
G}_{0{\bf p}}^<)_{22}-{\rm Im} (\hat{\rm G}_{0{\bf p}}^<)_{11}]
\nonumber\\
&=&\sum_{{\bf p}{\bf k}}|V(|{\bf p}-{\bf k}|)|^2 \cos\phi_{\bf
k}\sin(\phi_{\bf p}-\phi_{\bf k}) {\rm Re}(\hat{\rm G}_{1{\bf
k}}^{r})_{12}[{\rm Im} (\hat{\rm G}_{0{\bf p}}^<)_{22}-{\rm Im}
(\hat{\rm G}_{0{\bf p}}^<)_{11}].
\end{eqnarray}
From relation (\ref{TT}) it can be seen that ${\cal I}_1=0$.

Next, we analyze the remaining component of the integral, ${\cal
I}_2\equiv \sum_{\bf p} \sin\phi_{\bf p} {\rm Im} (\hat
F_2)_{12}$,
\begin{eqnarray}
{\cal I}_2&=&\sum_{\bf p}\sin\phi_{\bf p}\left \{ {\rm Re}(\hat {\rm
G}_{1{\bf p}}^<)_{12}[{\rm Im} (\Sigma_{0{\bf p}}^r)_{11}+{\rm Im}
(\Sigma_{0{\bf p}}^r)_{22}]\right .\nonumber\\
&&\left.
 +{\rm Im}(\hat {\rm G}_{1{\bf
p}}^<)_{12}[{\rm Re} (\Sigma_{0{\bf p}}^r)_{11}+{\rm Re}
(\Sigma_{0{\bf p}}^r)_{22}] -(\hat\Sigma\leftrightarrow\hat{\rm
G}) \right \}.\label{B4}
\end{eqnarray}
Again substituting the self-energies into Eq.\,(\ref{B4}), this
integral can be simplified as
\begin{eqnarray}
{\cal I}_2&=&\sum_{{\bf p},{\bf k}}|V({\bf p}-{\bf k})|^2
\cos\phi_{\bf k}\sin(\phi_{\bf p}-\phi_{\bf k}){\rm Re}(\hat{\rm G}_{1{\bf p}}^<)_{12}
[{\rm Im} (\hat {\rm G}_{0{\bf k}}^r)_{11}+{\rm Im} (\hat {\rm G}_{0{\bf k}}^r)_{22}]
\label{SIT}\\
&&-\frac 12\sum_{{\bf p},{\bf k}}[\cos\phi_{\bf k}-\cos\phi_{\bf p}\cos(\phi_{\bf p}-\phi_{\bf k})]
[{\rm Im}(\hat {\rm G}_{0{\bf p}}^r)_{11}+{\rm Im}(\hat {\rm G}_{0{\bf p}}^r)_{22}]
[{\rm Im}(\hat {\rm G}_{1{\bf k}}^<)_{11}-{\rm Im}(\hat {\rm G}_{1{\bf k}}^<)_{22}].\nonumber
\end{eqnarray}
Here, we have used Eq.\,(\ref{TT}). In this, the vanishing of the
real parts of the diagonal elements of the distribution
function-{\it i.e.}, ${\rm Re}(\hat {\rm G}_{1{\bf
k}}^<)_{\mu\mu}=0$-has also been considered. Combining the terms
proportional to $\cos\phi_{\bf p}\cos(\phi_{\bf p}-\phi_{\bf k})$
in the second line of Eq.\,(\ref{SIT}) with a similar term in the
first line, we obtain
\begin{eqnarray}
{\cal I}_2&=&\frac 12 \sum_{{\bf p}}\cos\phi_{\bf p}\left \{[{\rm
Im}(\hat \Sigma_{1{\bf p}}^<)_{11} -{\rm Im}(\hat\Sigma_{1{\bf
p}}^<)_{22}]
[{\rm Im} (\hat {\rm G}_{0{\bf p}}^r)_{11}+{\rm Im}
(\hat {\rm G}_{0{\bf p}}^r)_{22}]\right .\nonumber\\
&& \left. -[{\rm Im}(\hat\Sigma_{0{\bf p}}^<)_{11}+ {\rm
Im}(\hat\Sigma_{0{\bf p}}^<)_{22}] [{\rm Im} (\hat {\rm G}_{1{\bf
p}}^r)_{11}-{\rm Im}(\hat {\rm G}_{1{\bf p}}^r)_{22}]\right
\}.\label{CAI}
\end{eqnarray}
Further, we note that in the self-consistent approximation, there
is a vanishing quantity ${\cal K}$ given by
\begin{equation}
{\cal K}\equiv \sum_{{\bf p}}\cos \phi_{\bf p}\left \{[{\rm Im}
(\hat {\rm G}_{0{\bf p}}^r)_{11}-{\rm Im} (\hat {\rm G}_{0{\bf
p}}^r)_{22}] [{\rm Im}(\hat \Sigma_{1{\bf p}}^<)_{11}+{\rm
Im}(\hat\Sigma_{1{\bf p}}^<)_{22}] -(\Sigma\leftrightarrow {\rm
G})\right \}.
\end{equation}
The vanishing of ${\cal K}=0$ can be shown by inserting the
explicit forms of the self-energies, Eq.\,(\ref{SER}), into ${\cal
K}$ and employing Eq.\,(\ref{TT}). Finally, adding ${\cal K}$ to
the right hand side of Eq.\,(\ref{CAI}), we find
\begin{eqnarray}
{\cal I}_2&=&\frac 12 \sum_{{\bf p}\mu}\cos\phi_{\bf p}(-1)^{\mu+1}\left \{{\rm Im}(\hat \Sigma_{1{\bf p}}^<)_{\mu\mu}
{\rm Im} (\hat {\rm G}_{0{\bf p}}^r)_{\mu\mu}
-{\rm Im}(\hat\Sigma_{0{\bf p}}^<)_{\mu\mu}
{\rm Im} (\hat {\rm G}_{1{\bf p}}^r)_{\mu\mu}\right \}\nonumber\\
&=&\sum_{\bf p}\frac {\cos\phi_{\bf p}}2
[{\rm Re} ({\hat F}_2)_{11}-{\rm Re} ({\hat F}_2)_{22}],
\end{eqnarray}
which is Eq.\,(\ref{SR1}).

In the following, we prove Eq.\,(\ref{SR2}). From
Eq.\,(\ref{IL3}), the involved integral, denoted as ${\cal I}_3$,
can be written as
\begin{eqnarray}
{\cal I}_3&=&\frac {i}{4}\sum_{{\bf p}\mu\nu}\int
\frac{d\omega}{2\pi}(-1)^{\nu}\cos\phi_{\bf p} e{\bf E}\cdot
\nabla_{\bf p}\left \{ (\hat {\rm G}^<_0)_{\nu\nu}
 \frac{\partial}{\partial \omega}{\rm Re}(\hat {\Sigma}^r_0)_{\nu\nu}-
 \left[\frac{\partial}{\partial \omega}(\hat {\rm G}^<_0)_{\nu\nu}\right ]
{\rm Re}(\hat {\Sigma}^r_0)_{\nu\nu}\right .
\nonumber\\
&& \left . -\left[\frac{\partial}{\partial \omega}{\rm Re}(\hat
{\rm G}^r_0)_{\nu\nu}\right ] (\hat {\Sigma}^<_0)_{\nu\nu}+ {\rm
Re}(\hat {\rm G}^r_0)_{\nu\nu}
 \frac{\partial}{\partial \omega}(\hat {\Sigma}^<_0)_{\nu\nu}\right ].
\end{eqnarray}
Here, the momentum integral of Eq.\,(\ref{IL3}) has been performed
by parts integration, and the contribution from $(\hat R_2)_{12}$
vanishes due to its total derivative nature with respect to
$\omega$. Each part of the integrand involves two functions $\hat
{\rm G}^{r,<}_0$ and $\hat {\Sigma}^{r,<}_0$. Further integration
by parts with respect to $\omega$ in terms of the form
$\left(\frac{\partial}{\partial \omega}\nabla_{\bf p} {\rm
A}\right) {\rm B}$ ([with ${\rm A}$ and ${\rm B}$ being $(\hat
{\rm G}^{r,<}_0)_{\mu\mu}$ and $(\hat {\Sigma}^{r,<}_0)_{\nu\nu}$,
respectively, or vice versa] results in the operators $\nabla_{\bf
p}$ and $\partial/\partial \omega$ acting individually on ${\rm
A}$ or ${\rm B}$-{\it i.e.}, $\left(\frac{\partial}{\partial
\omega} {\rm A}\right) \nabla_{\bf p} {\rm B}$. Thus, we obtain
\begin{eqnarray}
{\cal I}_3&=&\frac {i}{2}\sum_{{\bf p}\mu\nu}\int
\frac{d\omega}{2\pi}(-1)^{\nu}\cos\phi_{\bf p} e{\bf E}\cdot \left
[\nabla_{\bf p}(\hat {\rm G}^<_0)_{\nu\nu}
 \frac{\partial}{\partial \omega}{\rm Re}(\hat {\Sigma}^r_0)_{\nu\nu}-
 \frac{\partial}{\partial \omega}(\hat {\rm G}^<_0)_{\nu\nu}
\nabla_{\bf p}{\rm Re}(\hat {\Sigma}^r_0)_{\nu\nu}\right .
\nonumber\\
&& \left .
-\frac{\partial}{\partial \omega}{\rm Re}(\hat {\rm G}^r_0)_{\nu\nu}
\nabla_{\bf p}(\hat {\Sigma}^<_0)_{\nu\nu}+
\nabla_{\bf p}{\rm Re}(\hat {\rm G}^r_0)_{\nu\nu}
 \frac{\partial}{\partial \omega}(\hat {\Sigma}^<_0)_{\nu\nu}\right ].
\end{eqnarray}
Using Eq.\,(\ref{IL2}), Eq.\,(\ref{SR2}) can be derived immediately.
\end{widetext}

\end{document}